\documentclass[prl,superscriptaddress,twocolumn]{revtex4}
\usepackage{graphics}
\usepackage{amsmath}
\usepackage{amssymb}
\usepackage{ulem}
\usepackage{amsfonts}
\usepackage{epic,eepic,epsfig}
\usepackage[active]{srcltx}
\usepackage{dcolumn}
\usepackage{bm}
\usepackage{color}
\usepackage{graphicx}

\newcommand{\bea}{\begin{eqnarray}}
\newcommand{\ena}{\end{eqnarray}}

\begin{document}
\title{Geometry of random potentials:  Induction of two-dimensional gravity in quantum Hall plateau transitions}
\author{Riccardo Conti}
\affiliation{Grupo de F\'isica Matem\'atica da Universidade de Lisboa, Av. Prof. Gama Pinto 2, 1649-003 Lisboa, Portugal.}
\author{Hrant Topchyan}
\affiliation{Alikhanyan National Laboratory, Yerevan Physics Institute, Armenia}
\author{Roberto Tateo}
\affiliation{Dipartimento di Fisica, Universit\`a di Torino, and INFN, Sezione di Torino, Via P. Giuria 1, I-10125 Torino, Italy}
\author{Ara Sedrakyan}
\affiliation{Alikhanyan National Laboratory, Yerevan Physics Institute, Armenia}

\begin{abstract} 

Integer Quantum Hall plateau transitions are usually modeled by a system of non-interacting electrons moving in a random potential. The physics of the most relevant degrees of freedom, the edge states, is captured by a recently-proposed random network model, in which randomness is induced by a parameter-dependent modification of a regular network. In this paper we formulate a specific map from random potentials onto 2D discrete surfaces, which indicates that 2D gravity emerges in all quantum phase transitions characterized by the presence of edge states in a disordered environment. We also establish a connection between the parameter in the network model and the Fermi level in the random potential.
\end{abstract}

\date{\today}

\maketitle

\noindent
{\it  Introduction.}  The physics of plateau transitions in the Quantum Hall Effect (QHE) continues to be one of the most exciting research topic in modern condensed matter physics. Much of the current interest is motivated by the emergence of a similar type of physics in the context of topological insulators. The Quantum Hall plateau transition is in fact an example of a metal-insulator transition (see \cite{Huckestein-1995, Ketemann-2005} for a review) with the plateau region between the Landau Levels (LLs) corresponding to the insulating phase where all the bulk states are localized due to the external magnetic field. The transition is a disorder-induced localization/delocalization transition of Anderson type, characterized by a divergent localization length $\xi$ with critical exponent $\nu$.\\
Quantum Hall plateau transitions can be modeled by a system of non-interacting electrons moving in a 2D random potential (RP) $V({\bf r})$, with ${\bf r}=(x,y)$, characterized by a white-noise Gaussian distribution.
In the following, we shall consider RPs with a finite correlation length generated by Gaussian sources placed on a regular lattice, i.e.
\begin{equation}
\label{eq:RP}
V({\bf r}) = \sum_{i,j} W_{i,j} \exp{\left(-\frac{|{\bf r}-{\bf r}_{i,j}|^2}{2\sigma^2} \right)} \;,
\end{equation}
where $\sigma^2$ is the variance, ${\bf r}_{i,j}=(i,j)$ is the position vector of the generic source  and the coefficients $W_{i,j}$ are randomly chosen in $[-W,W]$, for some $W\in\mathbb{R}$.
\begin{figure}[ht]
\centerline{\includegraphics[width=90mm,angle=0]{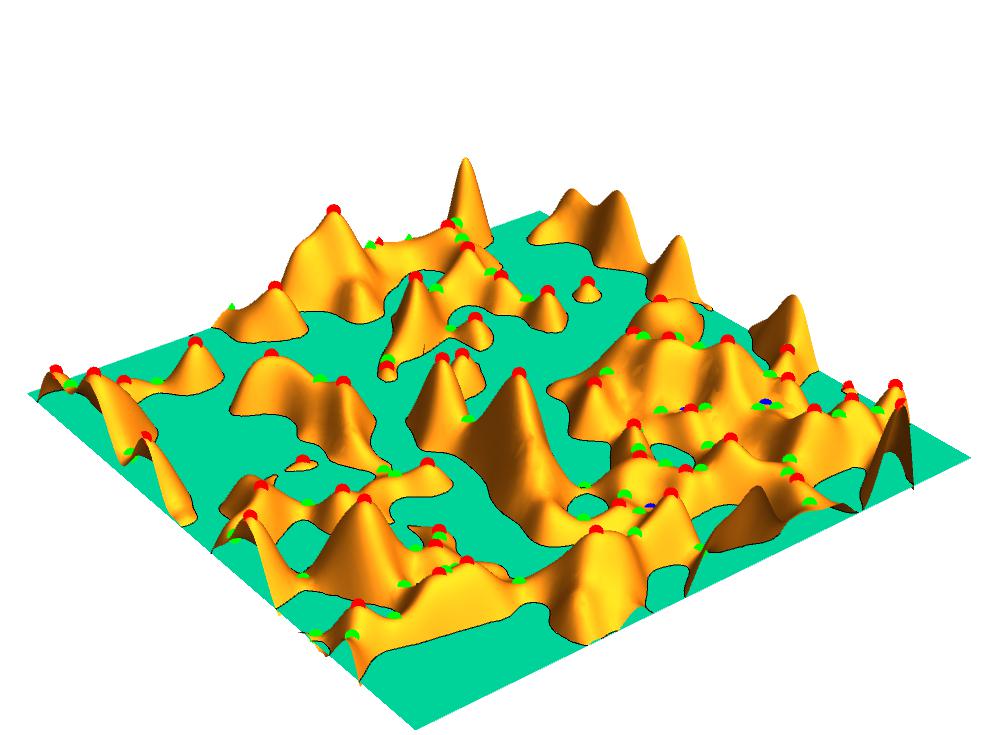}}
\caption{\small RP generated by $N=2500$ Gaussian sources ($W=1$ and $\sigma=\sqrt{2}$) placed on a torus. Points mark maxima (red), minima (blue) and saddle points (green). The plane represents the Fermi level ($c=0$). }
\label{1}
\end{figure}
In such RP landscape, electrons with energy smaller than the Fermi level $c$ are localized \cite{Abrahams-1979} due to the external magnetic field $B$ and their state corresponds semi-classically to an orbital motion with small radius $R_L \sim 1/B$. They fill the Fermi sea, which actually consists of a collection of lakes with characteristic 
size $l$, as displayed in FIG. \ref{1}. 
At the boundary of a lake, the orbital motion of (edge) electrons combines with the reflection due to the potential giving rise to a precession motion along equipotential lines. 

\begin{figure}[t]
	\includegraphics[width=50mm,angle=0,clip]
		{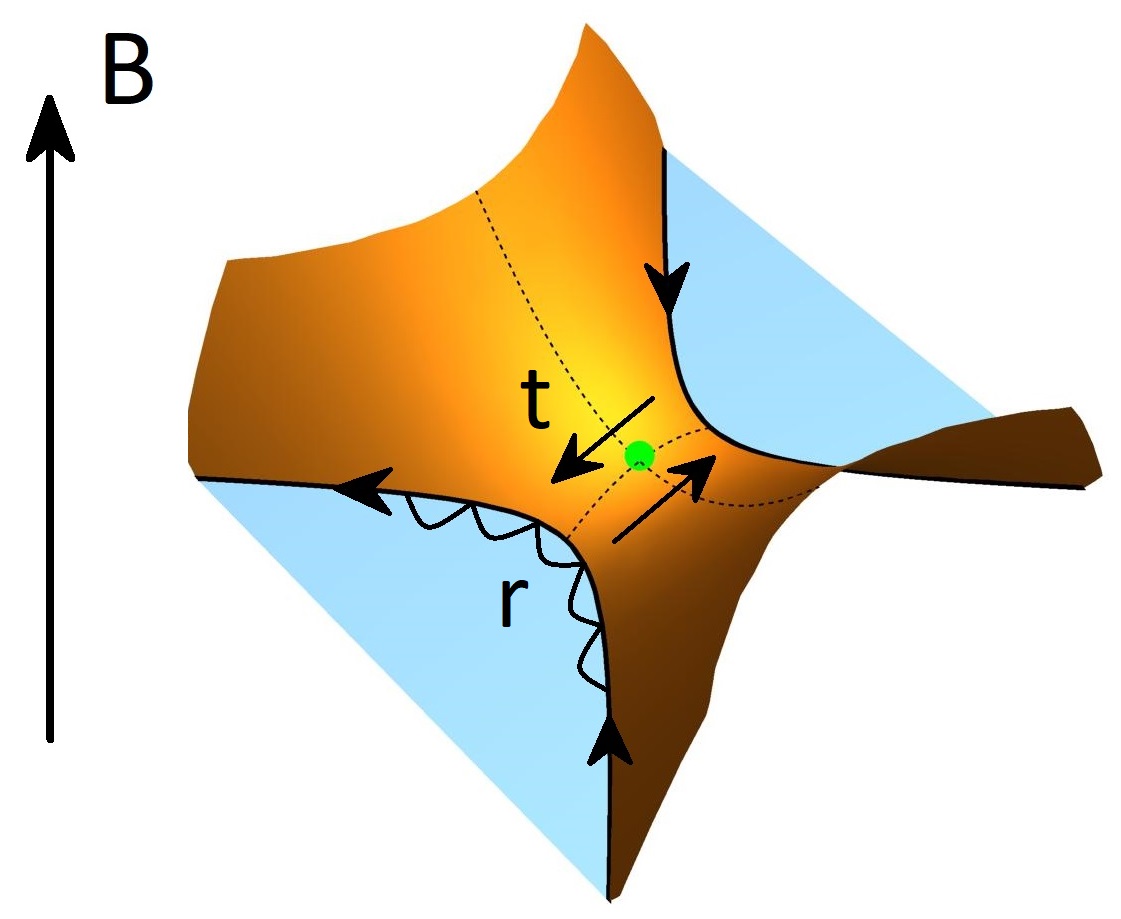}
	\caption{\small Neighborhood of a saddle point (green dot) separating two lakes (blue areas) in a RP.
	The cycloid represents the motion of edge states along the boundary of a lake. The parameters $r$ and $t$ denote the reflection and transmission probabilities, respectively, while $B$ is the magnetic field.} 
	\label{2}
\end{figure}

When an edge electron with energy $E>c$ approaches a saddle point, it may either tunnel through the potential barrier between the two neighbor lakes with probability \cite{FerHalp}
\begin{equation}
t^2\sim \frac{1}{1+e^\varepsilon} \;,\quad \varepsilon \propto (V-E) \;,
\end{equation}
or continue to move along the boundary of the same lake with probability $r^2=1-t^2$ (see FIG. \ref{2}). The presence of such quantum scattering nodes at saddle points enables electrons to reach arbitrary distances with a finite probability and is at the origin of the localization/delocalization transition.
Taking inspiration from this semi-classical picture, J. Chalker and P. Coddington (CC) \cite{Chalker-1988} formulated a network model of quantum scattering nodes based on a regular lattice that is meant to provide an effective description of the physics of edge states. Its generalization on a Kagome lattice was proposed in \cite{Sedrakyan-2020} and a similar network model for the Spin Quantum Hall Effect (SQHE) was studied in \cite{Kagalovski-1999,Gruzberg-1999}. Numerical investigations of the localization length $\xi$ around the critical point, i.e. $\xi \sim (t-t_{crit})^{-\nu}$ with $t_{crit}=1/\sqrt{2}$, resulted in $\nu =2.56\pm 0.62$ for a regular lattice \cite{Slevin-2009,Amado-2011,Beenakker-2011,Obuse-2012,Slevin-2012,Sedrakyan-2015} and $\nu= 2.658 \pm 0.046 $ for the Kagome lattice \cite{Sedrakyan-2020}. Both these values are not compatible with the experimental value $\nu=2.38 \pm 0.06$ measured for plateau transitions in the integer QHE \cite{Tsui-2005,Tsui-2009}. A possible solution to fix the discrepancy was put forward in \cite{Sedrakyan-2017, Sedrakyan-2019} by considering random networks (RNs), which should better account for the disorder present in a RP. 
The numerical estimate obtained in this framework $\nu=2.372 \pm 0.017$ \cite{Sedrakyan-2017, Sedrakyan-2019} confirms indeed a very good agreement with the experimental result.
 In fact, randomness generates -- in the continuum limit -- fluctuations of the background metric \cite{Sedrakyan-2017, Sedrakyan-2019}, namely 2D quantum gravity, that are responsibile for the change of the critical exponents in network models, similarly to what was established by \cite{KPZ} in the context of minimal models of statistical mechanics. The primary objective of this paper is to show that 2D gravity is indeed emerging from the RPs framework, by establishing a precise correspondence between RNs and RPs.
Notice that quantum gravity is also involved in the understanding of Fractional QHE \cite{Haldane-2011, Wiegmann-2015} revealing the physics of Laughlin wave-function. In that context, the interaction between fermions is responsible for the emergence of gravity in the bulk. Instead, in the present paper gravity is related to the 1+1 dimensional edge states, which originates from the RP. 

\vspace{0.3cm}

\begin{figure}[ht]
\centerline{\includegraphics[width=80mm,angle=0,clip]{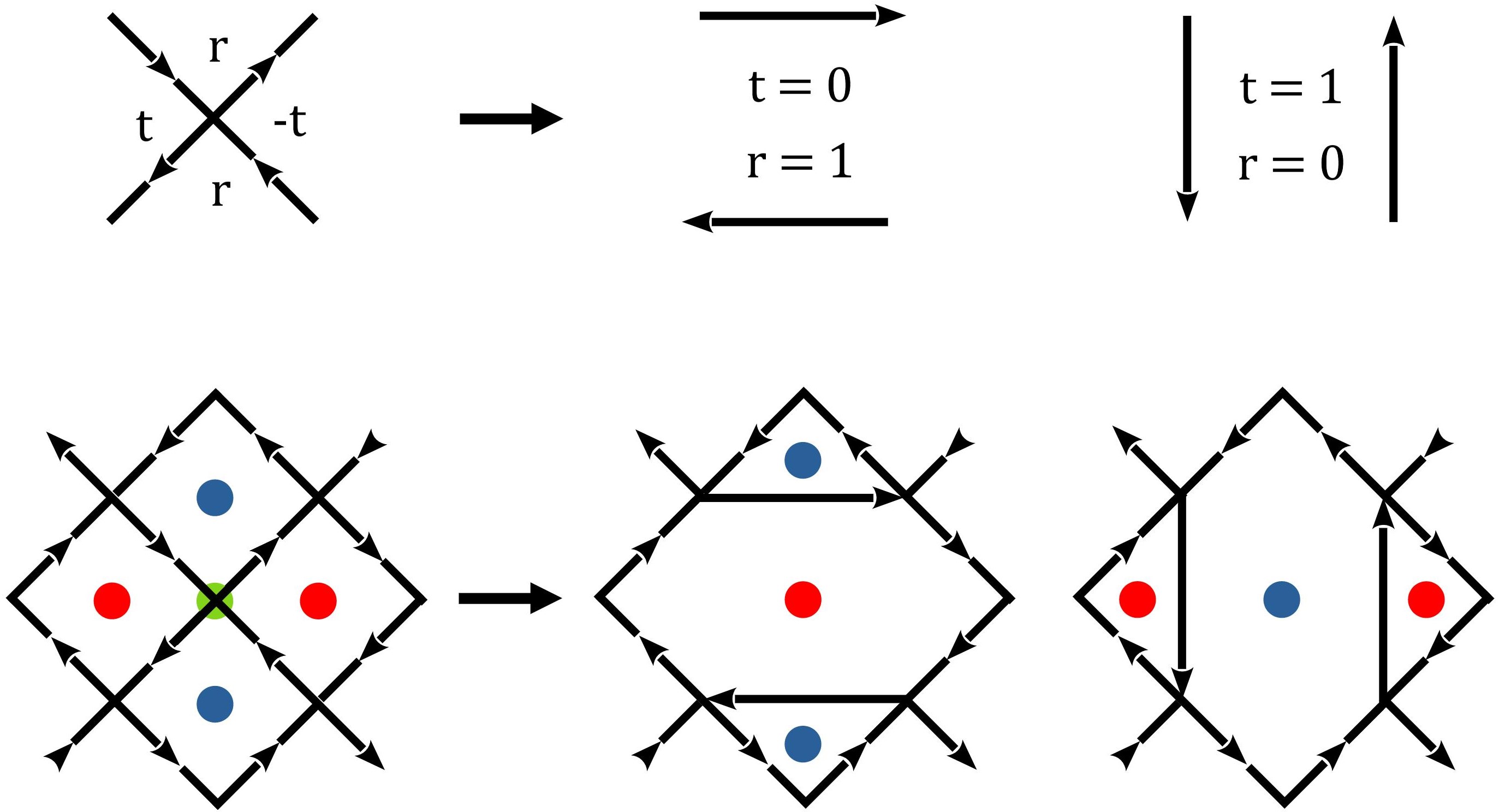}}
\caption{\small
Top: ``opening" of a scattering node in the horizontal and vertical directions. 
Bottom: result of the extreme replacement on the network. Red, blue and green points mark maxima, minima and saddle points in the corresponding RP framework.}
\label{3}
\end{figure}

\noindent
{\it  Network models with geometric disorder.} Let us briefly review the construction of RNs proposed in \cite{Sedrakyan-2017,Sedrakyan-2019}. Starting from a regular CC network, randomness is generated by making an extreme replacement, which consists in ``opening" a scattering node in the horizontal (vertical) direction with probability $p_0$ $(p_1)$ setting $t=0$ $(t=1)$ (see FIG. \ref{3}), or leaving it unchanged with probability $1-p_0-p_1$. In the following, we shall set $p_n=p_0=p_1$ to maintain statistical isotropy  \cite{Sedrakyan-2017,Sedrakyan-2019}. In the RP picture, the scattering node represents a saddle point and the four squares surrounding it correspond to an alternate sequence of maxima and minima (see FIG. \ref{3}). After the extreme replacement, the scattering node becomes an hexagon containing a maximum (minimum) and two adjacent triangles both containing a minimum (maximum), as depicted in FIG. \ref{3}.
Thus, starting from a regular network where all the faces are quadrangles and randomly making the extreme replacement with probability $p_n$, a polygonal tiling of the plane is obtained.
In \cite{Sedrakyan-2019} it was shown that in this type of RNs the critical index $\nu$ has a non-trivial dependence on the replacement probability $p_n$, with a critical line for $p_n\in [0,1/2]$. The best agreement with the experimental value of $\nu$ in the integer QHE is found for $( p_n, \nu(p_n))=(1/3,2.372 \pm 0.017) $.

A natural question addressed in the present paper concerns the physical interpretation of the parameter $p_n$ within the RP model. 

\vspace{0.2cm}

\begin{figure}[ht]
\centerline{\includegraphics[width=65mm,angle=0,clip]
		{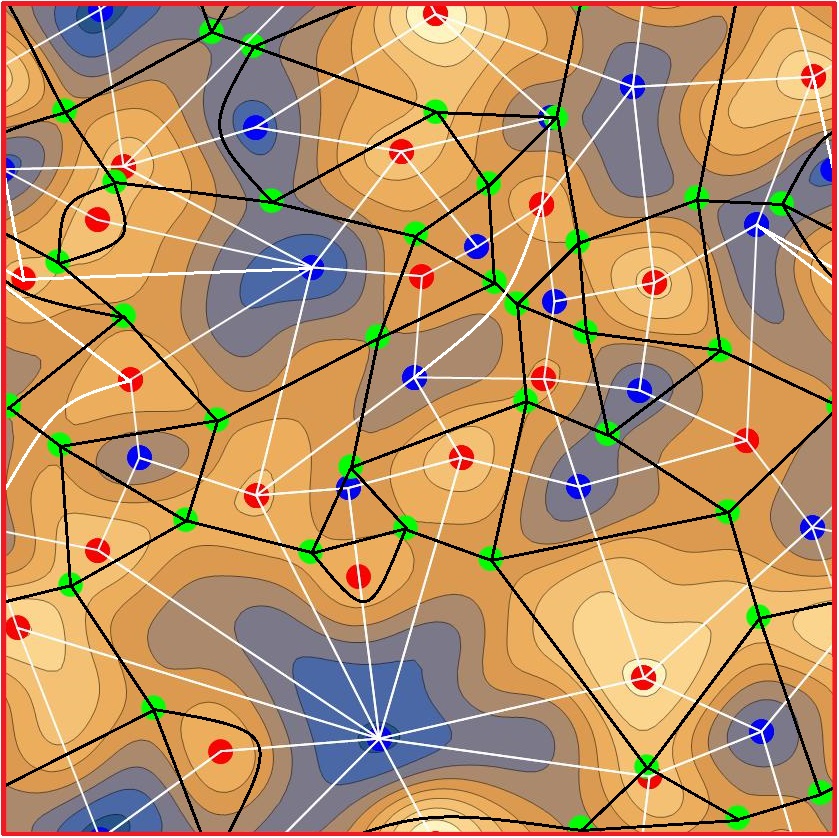}}
	\caption{\small Topography of a RP generated by $N=900$ Gaussian sources  ($W=1$ and $\sigma=\sqrt{2}$) placed on a torus. Points mark maxima (red), minima (blue) and saddle points (green). White and black lines are the edges of $S$ and $S^*$, respectively.}
	\label{4} 
\end{figure}

\noindent
{\it  Random potentials and discrete surfaces.} The RP \eqref{eq:RP} corresponds to a 2D smooth surface characterized by $N_{max}$ maxima, $N_{min}$ minima, $N_{sp}$ saddle points (see FIG. \ref{1}) and with Euler characteristics \cite{Milnor}
\begin{equation}
\label{eq:chi}
\chi= N_{min}+N_{max}-N_{sp}\;.
\end{equation}
Connecting maxima and minima according to the gradient of $V({\bf r})$ leads to a unique quadrangulation of the surface: a 2D discrete surface $S$ made of $v=N_{max}+N_{min}$ vertices, $e$ edges and $f=N_{sp}$ quadrangular faces (see FIG. \ref{4}). Denoting by $n_i$ the connectivity of the $i-$th vertex, i.e. the number of edges connected to it, the Euler characteristics $\chi=v-e+f$ of $S$ can be written as
\begin{equation}
\label{eq:chiR}
2\pi\chi = \sum_{i=1}^v R(n_i) \;,\quad R(n) = \frac{\pi}{2}(4-n) \;,
\end{equation}
where, according to Gauss-Bonnet theorem, $R(n)$ can be interpreted as the discrete Gaussian curvature associated to each vertex of $S$. Equation \eqref{eq:chiR} follows from $e=2f=\frac{1}{2}\sum_{i=1}^v n_i$, which implies
$\chi=v-e+f=v-f=\frac{1}{4}\sum_{i=1}^v (4-n_i)$.\\
By construction, each face of $S$ contains exactly one saddle point. Therefore, connecting saddle points belonging to nearest neighbor faces of $S$ results in a dual 2D discrete surface $S^*$. The latter surface is made of $v^*=f$ vertices with connectivity $4$, $e^*$ edges and $f^*=v$ polygonal faces of size $n$, where $n$ is the connectivity of the vertex of $S$ lying within each face of $S^*$ (see FIG. \ref{4}). By duality, each polygonal face of $S^*$ carries a discrete Gaussian curvature $R(n)$ and brings a local contribution to $2\pi\chi$, as described by eq. \eqref{eq:chiR}.
Hence, a RP is associated to a pair ($S,S^*$) of 2D discrete surfaces, which correspond to network models where the discrete Gaussian curvature of the surfaces is encoded either in the connectivity of the sites or in the number of sides of the polygons.
In the following, the symbols $S$ or $S^*$ will stand for both the discrete surfaces and the corresponding networks.
\begin{figure}[ht]
	\centerline{\includegraphics[width=65mm,angle=0,clip]
		{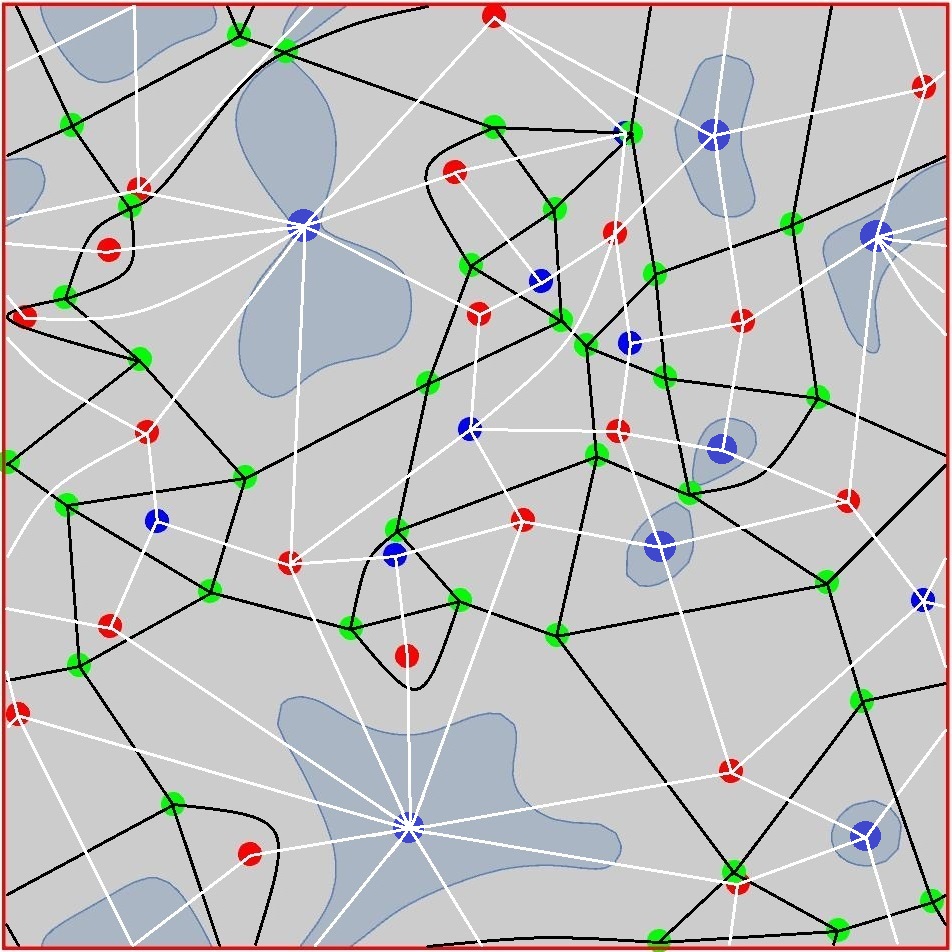}}
	\caption{\small Networks associated to the truncated discrete surfaces $S_c$ and $S_c^*$, obtained from the RP displayed in FIG. \ref{4} with $c=-1.2$. White and black lines are the links of $S_c$ and $S_c^*$, respectively. The areas highlighted in light blue indicate the regions under the Fermi level.}
	\label{5}
\end{figure}
%

\vspace{0.2cm}

\begin{figure}[ht]
	\centerline{\includegraphics[width=70mm,angle=0,clip]
		{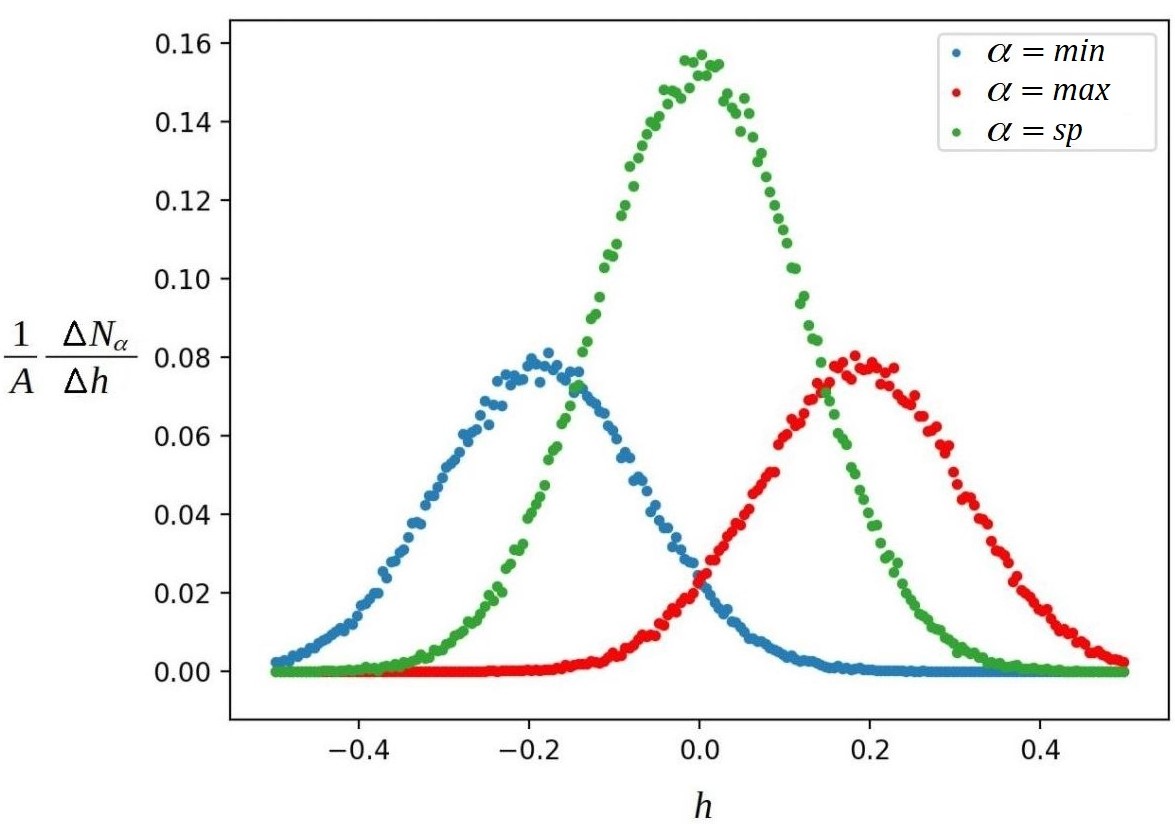}}
	\caption{\small Number of maxima $(\Delta N_{max})$, minima $(\Delta N_{min})$ and saddle points $(\Delta N_{sp})$ in the height range $[h,h+\Delta h]$, with $\Delta h=1/200$, divided by the area $A$ of the lattice. The statistical sample consists of $m=45$ simulations with $L=300$, i.e. $A=mL^2$, $W=1/10$ and $\sigma=\sqrt{2}$.} 
	\label{6}
\end{figure}

\noindent
{\it  Random potentials vs. Random networks.} 
The purpose of this section is to establish a correspondence between RPs and RNs in the case of a torus geometry. Consider a RP generated by $N=L^2$ Gaussian sources evenly distributed on a regular square lattice of size $L$ with unit spacing and doubly periodic boundary conditions. Let ${\bf r}_{i,j}= (i\mod{(L)},j\mod{(L)})$ be the position vector of the generic source on the lattice. Then, the RP at the generic point ${\bf r} = (x,y)\in [1,L]\times[1,L]$ is
\begin{equation}
\label{eq:RPtorus}
V({\bf r}) = \sum_{i,j=1}^L\sum_{{\bf n}\in\mathbb{Z}^2} W_{i,j} \exp{\left( -\frac{|{\bf r}-{\bf r}_{i,j} + {\bf n}L|^2}{2\sigma^2}\right)} \;,
\end{equation}
where the range of the summation index ${\bf n}=(n_x,n_y)$ is restricted to $\lbrace -1,0,1\rbrace\times\lbrace-1,0,1\rbrace$ in the numerical simulation. Equation \eqref{eq:chi} implies that $N_{max}+N_{min}=N_{sp}$, since $\chi=0$. In FIG. \ref{6}, the distributions of critical points per unit height $h$ of the potential are reported. The statistical sample consists of $m=45$ simulations with $L=300$, $W=1/10$ and $\sigma=\sqrt{2}$. Since at finite $W$ and $\sigma$ the potential $V({\bf r})$  is bounded, these distributions are defined on a finite support, also in the limit $L \rightarrow \infty$. However, in the case under consideration, they are well approximated by Gaussian distributions with expectation values $\mu_{max}=-\mu_{min}=0.187$, $\mu_{sp}=0$ and standard deviations $\sigma_{max}=\sigma_{min}=\sigma_{sp}=0.119$.
Following the procedure described in the previous section, a discrete surface $S$ or equivalently $S^*$ can be uniquely associated to the RP (see FIG. \ref{4}). The introduction of a Fermi level $c$ produces a truncated surface $S_c$ in which the vertices lying below $c$ and belonging to the same lake are replaced with a single vertex, as displayed in FIG. \ref{5}. 
This operation is indeed physically meaningful since the scattering of edge states is not affected by bulk electrons. Therefore, a change in the Fermi level induces a flow within the space of discrete surfaces parametrized by $c$.

The removal of sites due to the truncation generates polygonal faces in $S^*_c$ with different sizes compared to those of $S^*$ (see FIG. \ref{5}).
The net effect of this procedure is reminiscent of that induced in the CC network by the surgery defined in \cite{Sedrakyan-2017,Sedrakyan-2019} and leading to RNs. For this reason, we expect the replacement probability $p_n$ of RNs to be somehow related to the Fermi level in RPs.
\begin{figure}[t]
	\centerline{\includegraphics[width=80mm,angle=0,clip]
		{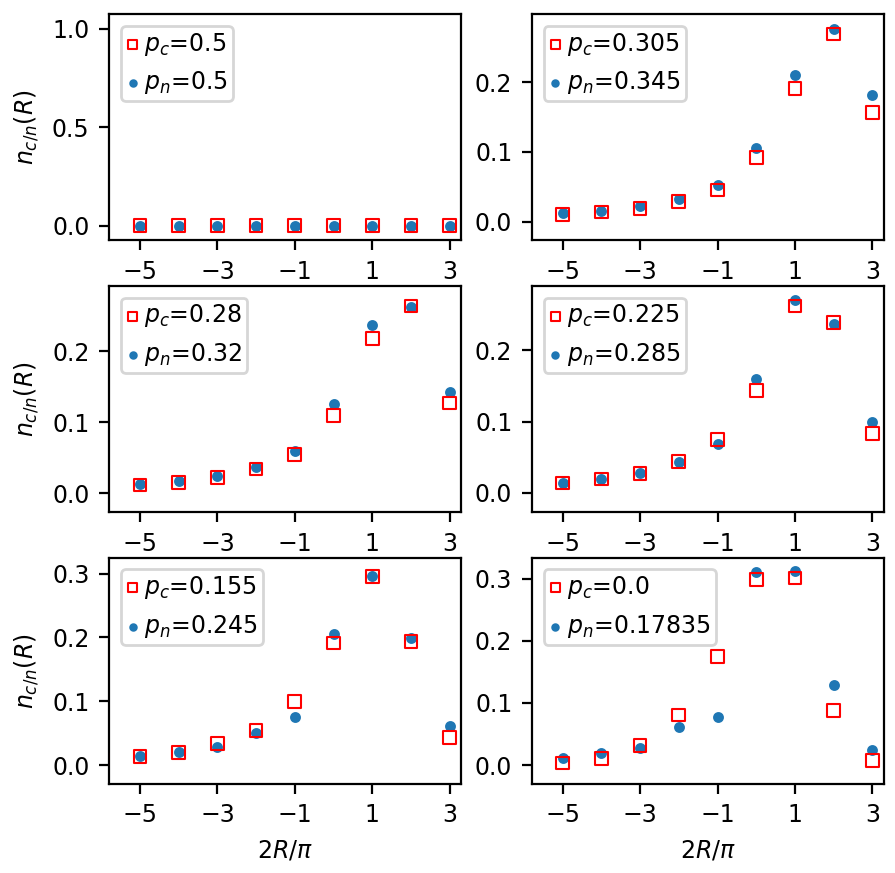}}
	\caption{\small Curvature distributions for the RN (blue dots) and the dual network $S_c^*$ (red squares) for various values of the parameters $p_n$ and $p_c$ which minimize the SSE.}
	\label{7}
\end{figure}
\begin{figure}[t]
	\centerline{\includegraphics[width=60mm,angle=0,clip]
		{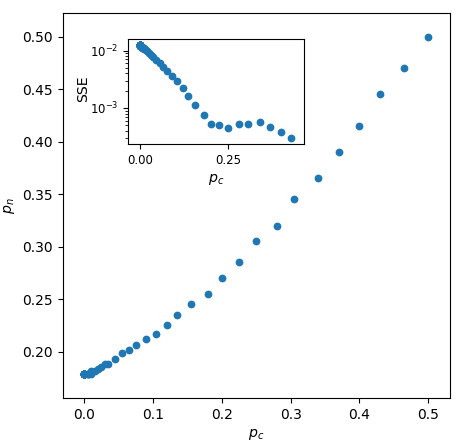}}
	\caption{\small 
Correspondence between the replacement probability $p_n$ and $p_c$ obtained searching for the best match between the two curvature distributions. The inset plot gives the estimated SSE as a function of $p_c$ 
in logarithmic scale.}
\label{8}
\end{figure}
However, for the purpose of comparing these two models, it is first necessary to restore the particle-hole duality in the RP framework because the RNs, which are described in the continuum limit by a Dirac fermion theory \cite{Sedrakyan-Bremen-2008}, possess it. To this aim, the range of energies accessible to an edge fermion in the RP should be restricted from $[c,+\infty]$ to the symmetric interval $I_c=[-|c|,|c|]$. We shall refer to the complementary interval $\bar I_c=[-\infty,-|c|]\cup[|c|,+\infty]$ as the {\it non-valid region}.
Next, notice that the replacement probability $p_n$ is equivalent -- at large network size -- to half the ratio of the number of removed scattering nodes to the total number 
of nodes. Therefore, the quantity
\begin{equation}
p_c=\frac{1}{2}\;\frac{\#\,\text{saddle points}\in\bar I_c}{\#\,\text{saddle points}\in (I_c\cup \bar I_c)} \;,
\end{equation}
appears to be the most appropriate parameter of the RP to be put in relation to $p_n$. 
Since the distribution of saddle points per unit height $h$ is approximately Gaussian (see FIG. \ref{6}), the  parameter $p_c$ can be related to the Fermi level via the complementary error function, $p_c \simeq \frac{1}{2}\mathbf{erfc}\left(|c|/(\sqrt{2}\sigma_{sp})\right)$.

To find the relation between $p_n$ and $p_c$, we consider the distribution  of discrete Gaussian curvatures $R$ of the polygons tiling both the RN and the dual network $S_c^*$ for several values of $p_n$ and $p_c$, respectively. The criterion adopted for the association between $p_n$ and $p_c$ is the minimization of the sum of squared errors, 
\begin{equation}
SSE=\sum_{m\geq 1} \left(n_{n}(R(m))-n_{c}(R(m))\right)^2 \;,
\end{equation}
where $n_{n}(R)$ and $n_{c}(R)$ denote the number of polygons with curvature $R$ divided by the total number of polygons in the RN and in $S_c^*$, respectively, with $R(m)$ as in eq. (\ref{eq:chiR}).
In FIG. \ref{7}, curvature distributions in both the RN and $S_c^*$ are compared for some values of $p_n$ and $p_c$ that minimize the SSE. The statistical samples consist of more than $50$ RN simulations on a $100\times 1000$ network for each value of $p_n \in [0,1/2]$ and $45$ RP simulations on a square lattice of size $L=300$ for each value of $c\in[0,1/2]$. A good agreement between the two models is obtained for a suitable correspondence $p_n\leftrightarrow p_c$, as reported in FIG. \ref{8}. 
We see that for $p_c \gtrsim 0.35 $ the relation $p_n(p_c)$ is approximately $p_n = p_c$, while for smaller values of $p_c$ the curve is deviating from the linear behavior ending at $p_n(0) \simeq 0.178$.
The origin of this deviation is related to the fact that $p_n=0$ corresponds to a regular network which can be associated to a periodic potential, while the RP is intrinsically disordered for any value of $p_c$. Periodic potentials have zero measure in the space of all RPs, therefore it is not surprising that the distribution of curvatures in the RN is less sensitive to variations of $p_n$ around $p_{n}=0$. Similar considerations might also justify the discrepancy between $n_n(R)$ and $n_c(R)$ that can be observed in the bottom right plot of FIG. \ref{7}. We shall leave a more systematic study of this issue to the future.

\bigskip

\noindent 
{\it  Conclusions.}
There are strong evidences that the field-theory description of plateau transitions corresponds to a model of fermions interacting with random gauge and scalar potentials and also with structurally-disordered geometry. Indicating that, in the scaling limit, localization transitions of this type are correctly described by matter fields coupled to 2D quantum gravity. Starting from a random potential model, we have explicitly constructed a map onto the 2D disordered graphs $S_c$ and $S^*_c$ depending on the Fermi-level. Thus, observing the appearance of the basic ingredient of random network models \cite{Sedrakyan-2017, Sedrakyan-2019} for Quantum Hall plateau transitions and giving an interpretation of the replacement probability in term of the Fermi energy. $S_c$ and $S^*_c$, being quadrangular and polygonal tilings of the plane, have a straightforward interpretation as discrete random surfaces, explicitly showing the emergence of 2D gravity. As discussed also in \cite{Sedrakyan-2017}, the notion of functional measure of random surfaces remains an open problem. From the current analysis, it appears that the distribution of Gaussian curvatures on the random surface associated with the random potential coincides with the corresponding distribution in the random network model, suggesting that the functional measure of random surfaces can be defined in terms of the measure of random potentials.
In conclusion, we revealed a deep link between random potentials in Anderson localization problem and 2D curved surfaces, where the edge states responsible for plateau transitions live. 
\bigskip

\noindent
{\it Acknowledgments.} A.S. acknowledge University of Turin and INFN for hospitality and ARC grant 18T-1C153 for financial support. This work was also partially supported by the INFN project SFT and by the FCT Project PTDC/MAT-PUR/30234/2017 ``Irregular connections on algebraic curves and Quantum Field Theory". R.C. is supported by the FCT Investigator grant IF/00069/2015 ``A mathematical framework for the ODE/IM correspondence''.

\bibliographystyle{apsrev4-2}

\begin{thebibliography}{99}
	
\bibitem{Huckestein-1995} B. Huckestein, Rev. Mod. Phys. 67, 357 (1995).

\bibitem{Ketemann-2005} B. Kramer, T. Ohtsuki and S. Kettemann
Phys. Rep. 417, 211 (2005).

\bibitem{Abrahams-1979} 
E. Abrahams, P. W. Anderson, D. C. Licciardello, and T. V.
Ramakrishnan, Phys. Rev. Lett. 42, 673 (1979).
	
\bibitem{Chalker-1988}
J. T. Chalker and P. D. Coddington, J. Phys. C 21, 2665 (1988).

\bibitem{Sedrakyan-2020} N. Charles, I. Gruzberg, A. Kl\"umper, W. Nudding, and A. Sedrakyan, ArXiv:2003.08167.

\bibitem{Kagalovski-1999} V. Kagalovsky, B. Horovitz, Y. Avishai, and J. T. Chalker, Phys. Rev. Lett. 82, 3516 (1999).

\bibitem{Gruzberg-1999} I. A. Gruzberg, A. W. W. Ludwig, and N. Read, Phys. Rev. Lett.82, 4524 (1999).

\bibitem{Slevin-2009} K. Slevin and T. Ohtsuki, Phys. Rev. B 80, 041304(R) (2009).

\bibitem{Amado-2011} M. Amado, A. V. Malyshev, A. Sedrakyan, and F. Domínguez-
Adame, Phys. Rev. Lett. 107, 066402 (2011).

\bibitem{Obuse-2012} H. Obuse, I. A. Gruzberg, and F. Evers, Phys. Rev. Lett. 109, 206804 (2012).

\bibitem{Slevin-2012} K. Slevin and T. Ohtsuki, Int. J. Mod. Phys. Conf. Ser. 11, 60 (2012).

\bibitem{Sedrakyan-2015} W. Nuding, A. Kl\"umper, and A. Sedrakyan, Phys. Rev. B 91, 115107 (2015).

\bibitem{Beenakker-2011} J. P. Dahlhaus, J. M. Edge, J. Tworzydło, and C. W. J.Beenakker, Phys. Rev. B 84, 115133 (2011).

\bibitem{Tsui-2005} W. Li, G. A. Cs\v{a}thy, D. C. Tsui, L. N. Pfeiffer, and K. W. West, Phys. Rev. Lett. 94, 206807 (2005).

\bibitem{Tsui-2009} W. Li, C. L. Vicente, J. S. Xia, W. Pan, D. C. Tsui, L. N.
Pfeiffer, and K. W. West, Phys. Rev. Lett. 102, 216801 (2009).


\bibitem{Sedrakyan-2017} I. A. Gruzberg, A. Kl\"umper, W. Nuding, and A. Sedrakyan, Phys. Rev. B 95, 125414 (2017).

\bibitem{Sedrakyan-2019} A. Kl\"umper, W. Nuding, and A. Sedrakyan, Phys. Rev. B 100, 140201(R) (2019).

\bibitem{KPZ} V. G. Knizhnik, A. M. Polyakov, and A. B. Zamolodchikov,
Mod. Phys. Lett. A 03, 819 (1988).

\bibitem{FerHalp} H.A. Fertig and B.I. Halperin, Phys. Rev. B 36, 7969 (1987).

\bibitem{Sedrakyan-Bremen-2008} A. Sedrakyan, Continuum limit of Chalker and Coddington
network model, Talk at Bremen conference, July, 2008.



\bibitem{Haldane-2011} F. D.M. Haldane, Phys.Rev.Lett. 107, 116801 (2011).

\bibitem{Wiegmann-2015} T. Can, M. Laskin, P. B. Wiegmann, Annals of Physics 362 (2015) 752–794.

\bibitem{Milnor} J. Milnor, Morse Theory-Princeton University Press, Pinceton, New Jersey 1963.



\end{thebibliography}

\end{document}